\documentclass[conference,10pt,a4paper]{IEEEtran}
\IEEEoverridecommandlockouts
\usepackage{cite}
\usepackage{amsmath,amssymb,amsfonts}
\usepackage{algorithmic}
\usepackage{graphicx}
\usepackage{textcomp}
\usepackage{xcolor}
\def\BibTeX{{\rm B\kern-.05em{\sc i\kern-.025em b}\kern-.08em
    T\kern-.1667em\lower.7ex\hbox{E}\kern-.125emX}}

\usepackage{pgfplots}
\pgfplotsset{compat=newest}
\pgfplotsset{plot coordinates/math parser=false} 
\usepgfplotslibrary{groupplots}
\usepackage{tikz}
\usetikzlibrary{matrix}
\usetikzlibrary{positioning}
\tikzset{
	position/.style args={#1:#2 from #3}{
		at=(#3.#1), anchor=#1+180, shift=(#1:#2)
	}
}
\newlength\figureheight
\newlength\figurewidth 

\usepackage{hhline}
\usepackage{subfig}

\definecolor{mycolorred}{rgb}{0.85000,0.32500,0.09800}%
\definecolor{mycolorblue}{rgb}{0.00000,0.44700,0.74100}%
\definecolor{mycolorpurple}{rgb}{0.49400,0.18400,0.55600}%
\definecolor{mycolorgreen}{rgb}{0.46600,0.67400,0.18800}%
\definecolor{mycolorblack}{rgb}{0.00,0.00,0.00}%
\definecolor{mycoloryellow}{rgb}{1,0.65,0}%

\begin{document}

\title{Scalable Lossless Coding of Dynamic Medical CT Data Using Motion Compensated Wavelet Lifting with Denoised Prediction and Update}

\author{\IEEEauthorblockN{Daniela Lanz, Franz Schilling, and Andr\'{e} Kaup}
	\IEEEauthorblockA{Multimedia Communications and Signal Processing\\
		Friedrich-Alexander University Erlangen-N\"{u}rnberg (FAU)\\
		Cauerstr. 7, 91058 Erlangen, Germany\\
		\{Daniela.Lanz,Franz.Schilling,Andre.Kaup\}@FAU.de}}

\maketitle

\begin{abstract}
	Professional applications like telemedicine often require scalable lossless coding of sensitive data. \mbox{3-D subband} coding has turned out to offer good compression results for dynamic CT data and additionally provides a scalable representation in terms of low- and highpass subbands. To improve the visual quality of the lowpass subband, motion compensation can be incorporated into the lifting structure, but leads to inferior compression results at the same time. Prior work has shown that a denoising filter in the update step can improve the compression ratio. In this paper, we present a new processing order of motion compensation and denoising in the update step and additionally introduce a second denoising filter in the prediction step. This allows for reducing the overall file size by up to $4.4\%$, while the visual quality of the lowpass subband is kept nearly constant.
\end{abstract}

\vspace{0.5ex}

\IEEEpeerreviewmaketitle

\vspace{-2ex}
\section{Introduction}
\begin{figure}
	\includegraphics[width=\columnwidth]{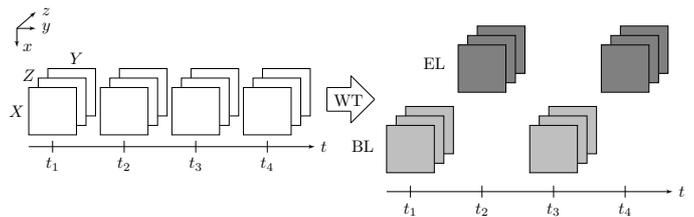}
	\caption{Considered scenario to achieve temporal scalability of a 3-D+$t$ volume using a wavelet transform in temporal direction.}
	\label{fig:data}
	\vspace{-3ex}
\end{figure}

Medical imaging, like Computed Tomography\,(CT), requires lossless compression due to the essential content. However, lossless compression leads to high bit rates, which is challenging for fast transmitting and archiving the data. Especially modern multi-slice CT scanners, which allow for recording volumes of the human body over a specific time, can generate up to $200$\,MB/s~\cite{Ohnesorge2007}. The structure of such a dynamic 3-D+$t$ volume can be seen on the left side of Fig.\,\ref{fig:data}. Here, it consists of $T{=}4$ temporally equidistant spatial \mbox{3-D} volumes of size $X{\times}Y{\times}Z$ and can be splitted into $Z{=}3$ temporal sequences. Then, every temporal sequence can be encoded using state-of-the-art compression techniques.

Established video compression standards, like the High Efficiency Video Coding\,(HEVC) standard~\cite{2016}, have mainly been developed for consumer applications and are specialized for lossy compression. In~\cite{Lanz2018}, it has been shown that for lossless compression of dynamic CT data, 3-D subband coding\,(SBC)~\cite{Karlsson1988} performs better than HEVC. Since 3-D SBC is based on discrete wavelet transforms\,(WT), it inherently provides scalability features. This allows for transmitting a base layer\,(BL) with coarser quality and an enhancement layer\,(EL) on demand, comprising the residual data to reconstruct the original sequence. In Fig.\,\ref{fig:data}, a WT in temporal direction is performed, decomposing every single temporal sequence into two layers, both comprising only half the frame rate.
This is very advantageous for telemedicine applications, like remote surgery systems or transmission of medical content for remote assessment over wireless networks with limited channel capacity. Incorporating motion compensation\,(MC) methods into the lifting structure of a WT is called motion compensated temporal filtering\,(MCTF)~\cite{334985} and increases the quality of the BL significantly, but results also in a lower compression efficiency. To overcome this drawback, wavelet lifting with denoised update\,(WLDU)~\cite{Lanz2018} was recently introduced.

To further improve the compression efficiency of dynamic CT data, in this paper we propose to change the processing order of MC and denoising in the update step of WLDU. Besides this, we introduce a second denoising operation in the prediction step, which leads to additional coding gains.
After a short recap of compensated wavelet lifting in Section~\ref{sec:sbc}, the two proposed modifications will be explained in detail in Section~\ref{sec:prop}. Simulation results are given in Section~\ref{sec:results}, while Section~\ref{sec:conclusion} concludes this work.

\section{Compensated Wavelet Lifting}
\label{sec:sbc}
\begin{figure}
	\centering
	\includegraphics[width=0.73\columnwidth]{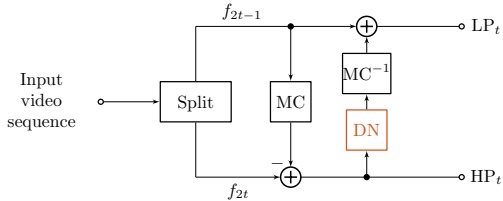}
	\caption{Block diagram of the motion compensated lifting structure containing the denoising filter (DN) of WLDU~\cite{Lanz2018}.}
	\label{fig:liftingMod}
	\vspace{-2ex}
\end{figure}

Discrete wavelet transforms can be implemented very efficiently by the so-called lifting structure~\cite{Sweldens1995}, which consists of three steps. In Fig.\,\ref{fig:liftingMod}, a block diagram of the lifting structure for a WT in temporal direction is shown. In the first step, named splitting, the input signal is decomposed into even- and odd-indexed frames $f_{2t}$ and $f_{2t-1}$. In the next step, called prediction step, the even frames are predicted from the odd frames by a prediction operator $\mathcal{P}$. After subtracting the predicted values from $f_{2t}$, the highpass\,(HP) coefficients $\text{HP}_t$ are achieved. The last step is called update step and filters the HP coefficients by an update operator $\mathcal{U}$. Then, the HP coefficients are added to the odd frames, resulting in the lowpass\,(LP) coefficients $\text{LP}_t$. Since the resulting LP subband is very similar to the original signal, it directly serves as the BL. In contrast, the HP subband comprises the residual data and therefore corresponds to the EL.

The lifting structure offers a very flexible framework and can be modified in multiple ways. To guarantee lossless reconstruction, rounding operators are applied to achieve integer-to-integer transforms~\cite{647983}. Additionally, MC methods can directly be incorporated into the lifting structure to avoid blurring and ghosting artifacts due to temporal displacements in the sequence. This is called MCTF and is achieved by realizing the prediction operator $\mathcal{P}$ by the warping operator $\mathcal{W}_{2t-1\rightarrow 2t}$. Accordingly, for the Haar wavelet filters, the HP coefficients are calculated by
\begin{equation}
	\text{HP}_t = f_{2t} - \lfloor \mathcal{W}_{2t-1\rightarrow 2t}(f_{2t-1})\rfloor.
\end{equation}
To achieve an equivalent WT, in the update step, the compensation has to be inverted. Consequently, the LP coefficients of the Haar transform are calculated by reversing the index of $\mathcal{W}$
\begin{equation}
	\text{LP}_t = f_{2t-1} + \lfloor \frac{1}{2} \mathcal{W}_{2t\rightarrow 2t-1}(\text{HP}_t)\rfloor.
\end{equation}
However, for noisy CT data, the required bit rate to encode the subbands will rise by applying MCTF. This is due to the fact that not only structural information is warped from one frame to the other, but also noise is warped back and forth. 

To overcome the noise problem, prefiltering of the even and odd frames can be applied. However, lossless reconstruction is not provided anymore and therefore, such schemes are not suitable for professional applications, in which lossless compression is claimed. 
One way to reduce this effect without compromising lossless reconstruction is to skip the update step entirely, which is called truncated WT. However, the resulting LP subband suffers not only from temporal aliasing, but loses also the property of being a good representative for both even- and odd-indexed frames. This makes the truncated WT useless for professional applications. 
Another way to reduce the increase of the noise variance in the LP subband and to preserve lossless reconstruction was recently investigated in~\cite{Lanz2018} and is called wavelet lifting with denoised update\,(WLDU). The basic scheme can be seen in Fig.\,\ref{fig:liftingMod} highlighted in red. It has been shown that the application of suitable filters for denoising\,(DN) in the update step leads to a rising compression ratio of the LP subband, while the visual quality of the LP subband can be preserved for being used in telemedicine applications. 

\section{Proposed Modifications}
\label{sec:prop}

\begin{figure}
	\centering
	\subfloat{\includegraphics[width=0.6\columnwidth]{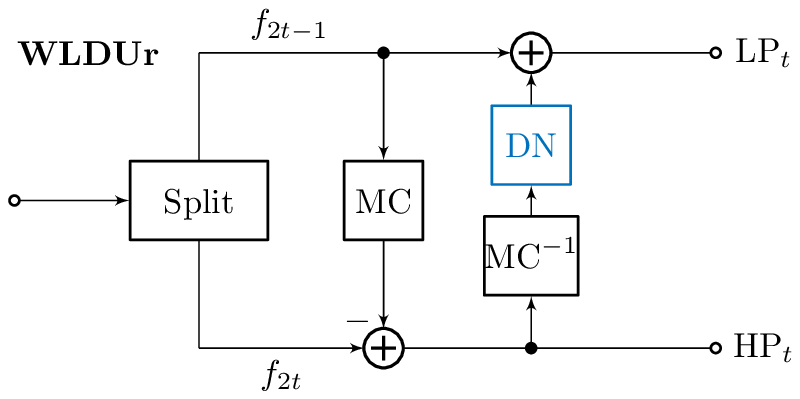}}\\ \vspace{-1.5ex}
	\subfloat{\includegraphics[width=0.6\columnwidth]{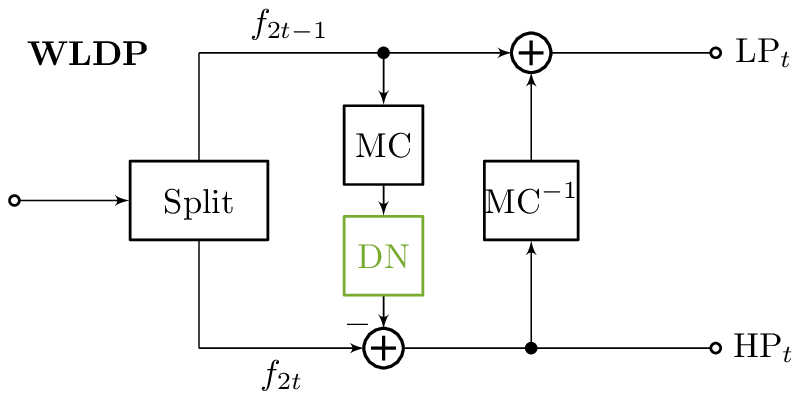}}\\ \vspace{-1.5ex}
	\subfloat{\includegraphics[width=0.6\columnwidth]{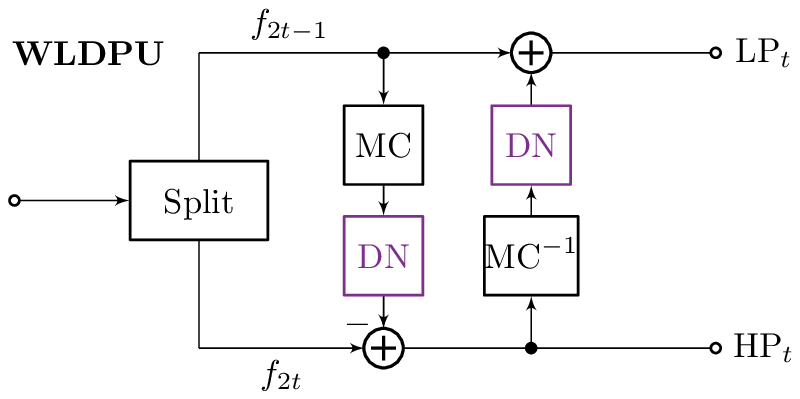}}
	\caption{Block diagrams of the different proposed setups. From top to bottom: WLDUr, WLDP, and WLDPU.}
	\label{fig:liftingMods}
	\vspace{-2ex}
\end{figure}

\begin{figure*}[t]
	\centering
	\resizebox{\textwidth}{!}{%
		\input{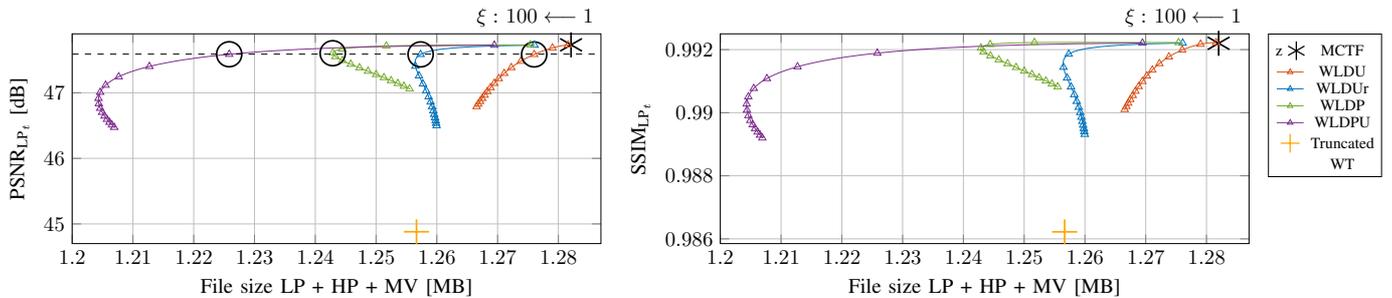}}
	\caption{$\text{PSNR}_{\text{LP}_t}$ (left) and $\text{SSIM}_{\text{LP}_t}$ (right) results, respectively, against the overall file size comparing WLDU, WLDUr, WLDP, and WLDPU. The arrow above the diagrams indicates the direction for increasing values of the filter strength $h{=}\xi{\cdot}\sigma_n^2$. For better presentation, only every eighth value for $h$ is plotted. Results are averaged over all 327 sequences.}
	\label{fig:val}
	\vspace{-2ex}
\end{figure*}

To further increase the compression ratio, we introduce two novel modifications in this paper, resulting in three different setups. First, we propose to change the processing order in the update step of WLDU as it is shown in Fig.\,\ref{fig:liftingMods} on the top. In the following, this setup will be called WLDU with reversed order\,(WLDUr). 
However, by applying DN in the update step, only the compression ratio of the LP subband is affected. The HP subband is not touched so far. Therefore, the second setup targets the prediction step, as can be seen in Fig.\,\ref{fig:liftingMods} in the middle, called wavelet lifting with denoised prediction\,(WLDP).
The third setup is given by the combination of WLDUr and WLDP, called wavelet lifting with denoised prediction and update\,(WLDPU) and can be seen in Fig.\,\ref{fig:liftingMods} on the bottom.

\subsection{WLDU with reversed order}
\label{subsec:order}
As already described above, MCTF reduces the appearance of ghosting artifacts. However, it also increases the noise variance $\sigma_n^2$ of the single subbands, leading to a higher overall entropy~\cite{lnt2012-10} and therefore to a higher number of bits required for coding the subbands.
To compensate this drawback, WLDU was introduced. By denoising the HP frames before warping them back to the time step of the odd frames, the undesired increase of $\sigma_n^2$ in the HP subband caused by the MC in the prediction step was reduced.
However, by inversely compensating the HP frames and adding them to the odd frames in the update step, the noise variance of the LP subband can be increased further due to the inverse compensation process. Therefore, we propose to warp the HP frames at first to the time step of the odd frames and to denoise them by a suitable filter afterwards. By applying WLDUr, erroneously warped objects and noisy structures from both warping operations can be suppressed more efficiently, leading to a better compression ratio of the LP frames.  

\subsection{Wavelet Lifting with Denoised Prediction}
\label{subsec:wldp}
The second proposed modification concerns the prediction step. So far, the DN filter of WLDU and WLDUr has only influenced the compression ratio of the resulting LP subband. However, for lossless reconstruction of the original sequence, the file size of the corresponding HP subband is also of interest. Therefore, we introduce a second DN filter in the prediction step similar to~\cite{faucris.122127984}, where in-loop denoising is performed on reference frames in the context of HEVC. 
According to them, in-loop prediction filters can be applied on the reference frames themselves or on the motion compensated frames. However, by performing DN on the reference frames, for every DN setup, including the specific filter type and the chosen filter strength, new motion vectors have to be calculated. 
This makes the whole framework very inflexible. Therefore, we decided to perform DN on the motion compensated frames. 

\subsection{Wavelet Lifting with Denoised Prediction and Update}
\label{subsec:wldpu}
Since WLDP influences also the performance of the denoising in the update step, the combination of both modifications is also considered. By forcing noise reduction in both lifting steps, we expect a further increase of the compression ratio.

\section{Simulation Results}
\label{sec:results}

\begin{table*}[t]	
	\begin{minipage}{0.74\textwidth}
		\caption*{TABLE I: Overall file size and $\text{PSNR}_{\text{LP}_t}$ for certain values of Fig.~\ref{fig:val} using BM3D filtering. Absolute and relative differences against MCTF are given. Negative numbers denote a smaller file size and a lower $\text{PSNR}_{\text{LP}_t}$ compared to MCTF. Results are averaged over all 327 sequences.}
		\setlength{\tabcolsep}{2pt}
		\resizebox{\textwidth}{!}{%
			\begin{tabular}{l|c|c|c|c||c|c}
				Method & Noise & File size & \multicolumn{2}{c||}{$\Delta$ to MCTF}   & $\text{PSNR}_{\text{LP}_t}$ {[}dB{]} & $\Delta$ to MCTF {[}dB{]} \\ \hhline{~|~|~|--||~|~}
				& parameter & LP + HP + MV {[}MB{]} & Absolute {[}MB{]} & Relative {[}$\%${]} &   &     \\ \hline
				MCTF                       &             & 1.2821 & -       & -       & 47.7322 & -     \\
				$\text{WLDU}$  & $\xi{=}16$  & 1.2760 & -0.0061 & -0.4758 & 47.5859 & -0.1463 \\
				\textcolor{mycolorblue}{$\text{WLDUr}$} & $\xi{=}\hphantom{1}8$  & 1.2573 & -0.0248 & -1.9343 & 47.5861 & -0.1461 \\
				\textcolor{mycolorgreen}{$\text{WLDP}$}  & $\xi{=}24$ & 1.2429 & -0.0392 & -3.0575 & 47.6041 & -0.1281 \\
				\textcolor{mycolorpurple}{$\text{WLDPU}$} & $\xi{=}\hphantom{1}8$  & 1.2258 & -0.0563 & -4.3912 & 47.5885 & -0.1437\\
				Truncated WT               &            & 1.2567 & -0.0254 & -1.9811 & 44.8806 & -2.8516     
			\end{tabular}}
			\label{tab:results}
		\end{minipage}\hspace{0.5cm}
		\begin{minipage}{0.22\textwidth}
			\centering
			\definecolor{mycolorred}{rgb}{0.85000,0.32500,0.09800}%
\definecolor{mycolorblue}{rgb}{0.00000,0.44700,0.74100}%
\definecolor{mycolorpurple}{rgb}{0.49400,0.18400,0.55600}%
\definecolor{mycolorgreen}{rgb}{0.46600,0.67400,0.18800}%
\definecolor{mycolorblack}{rgb}{0.00,0.00,0.00}%
\definecolor{mycoloryellow}{rgb}{1,0.65,0}%
\begin{tikzpicture}

\begin{groupplot}[group style={group name = group,group size=1 by 1, horizontal sep = 2cm, vertical sep=1.1cm}]

%
%
%

\nextgroupplot[%
width=2.8cm,
height=2.4cm,
scale only axis,
xtick={1,2,3,4},
xticklabels={\scriptsize{{WLDUr}},\scriptsize{{WLDP}},\scriptsize{{WLDPU}}},
ytick={1,2,3,4,5},
xmin=0.5,
xmax=3.5,
ylabel={Bit rate savings in [\%]},
ymajorgrids,
axis background/.style={fill=white},
title style={at={(1,1.1)},anchor=north east},
legend style={legend cell align=left,align=left,draw=white!15!black}
]

\addplot [color=mycolorblue,dashed,mark=triangle,mark options={thick,solid}]
table[row sep=crcr]{%
	1	1.217\\
	1	1.966\\
	1	2.521\\
};

\addplot [color=mycolorgreen,dashed,mark=triangle,mark options={thick,solid}]
table[row sep=crcr]{%
	2	2.075\\
	2	3.116\\
	2	4.14\\
};

\addplot [color=mycolorpurple,dashed,mark=triangle,mark options={thick,solid}]
table[row sep=crcr]{%
	3	3.389\\
	3	4.399\\
	3	5.085\\
};


\end{groupplot}
\end{tikzpicture}
			\vspace{-1ex}
			\caption*{{Fig.\,5: Range of relative bit rate savings for all proposed setups compared to MCTF.}}
			\label{fig:variances}
		\end{minipage}	
		\vspace{-2ex}
	\end{table*}
	\setcounter{figure}{5}
	
	All experiments are evaluated on two 3-D+t medical CT data sets\footnote{The CT volume data sets were kindly provided by Siemens Healthineers.}, both describing a beating heart and comprising $10$ time steps and a spatial resolution of $512\times512$ pixels at a bit depth of $12$ bits per sample. The data sets consist of $127$ and $200$ slices per time step, respectively, summing up to $327$ single temporal sequences. 
	The MC is realized by a mesh-based approach~\cite{8066388} using a grid size of $8$ pixels. The resulting LP and HP subbands are encoded without any loss using the wavelet-based volume coder JPEG\,2000~\cite{ITU-T-800}. Here, the \mbox{OpenJPEG}~\cite{openjpeg} implementation with four spatial decomposition steps in $xy$-direction is applied. The arising motion vectors\,(MV) are encoded using the QccPack library~\cite{fowler2000qccpack}. To evaluate the quality of the LP subband, we use the $\text{PSNR}_{\text{LP}_t}$ metric as well as the modified structural similarity index $\text{SSIM}_{\text{LP}_t}$, which were both introduced in~\cite{Lanz2018}. Both consider not only the similarity of the LP frames to the odd-indexed frames, but also the similarity of the LP frames to the even-indexed frames. 
	All results are averaged over all $327$ sequences.
	
	To compare our novel modifications to WLDU, we apply Block Matching and 3-D Filtering\,(BM3D)~\cite{Dabov2007} for denoising, which was already used in~\cite{Lanz2018} and has given the highest compression gains. 
	The corresponding filter parameters are also chosen identically. The denoising is evaluated for different values of the filter strength $h$. The filter strength $h$ is an input parameter of BM3D and is calculated by
	\begin{equation}
		h = \xi \cdot \sigma_n^2,
	\end{equation}
	where $\sigma_n^2$ describes the noise variance of the input image and $\xi$ denotes an arbitrary noise parameter which controls the strength of denoising in order to improve the compression efficiency. A higher value of $\xi$ leads to a lower noise variance of the output image and according to~\cite{lnt2012-10} to a better compression ratio. We examine the influence of $h$ by varying $\xi$ in a range of integer values from $1$ to $100$. $\sigma_n^2$ is estimated by a low-complexity algorithm proposed in~\cite{IMMERKAER1996300} at both encoder and decoder side.
		
	\subsection{Comparing WLDUr, WLDP, and WLDPU}
	
	In Fig.~\ref{fig:val}, all considered setups are evaluated for the BM3D filter. For better displaying, only every eighth value of $h$ is plotted. Additionally to the $\text{PSNR}_{\text{LP}_t}$ and rate results on the left side, the $\text{SSIM}_{\text{LP}_t}$ results are provided on the right side. The single curves for both metrices show the same behavior. Therefore, all further analyses are evaluated only by $\text{PSNR}_{\text{LP}_t}$.
	
	As mentioned in Section~\ref{sec:sbc}, the diagram shows that the truncated WT results in a very low $\text{PSNR}_{\text{LP}_t}$ value. However, the compression efficiency of the truncated WT is higher than for WLDU with a very strong filter strength of $h{=}100{\cdot}\sigma_n^2$. 
	Further, it can be observed that all proposed setups outperform the compression efficiency of WLDU. By applying WLDUr, the quality of the LP subband is kept at a very high level until a filter strength of $h{=}8{\cdot}\sigma_n^2$, afterwards $\text{PSNR}_{\text{LP}_t}$ drops very fast. However, the achieved compression of BM3D for this filter strength gives similar results as the truncated WT. 
	By using WLDP, further bit rate savings compared to WLDUr can be observed. Until a filter strength of $h{=}24\cdot\sigma_n^2$ the rate decreases significantly, while the quality in terms of $\text{PSNR}_{\text{LP}_t}$ is nearly kept constant. Afterwards the rate starts increasing and the $\text{PSNR}_{\text{LP}_t}$ value decreases simultaneously. 
	The combination of both proposed modifications, shown by the violet curve for WLDPU, merges the properties of WLDUr and WLDP. This means that additional coding gains are achieved due to the influence of the DN operations to both subbands. 
	
	For a fair comparison of all proposed setups, we draw a horizontal line at a constant level of $\text{PSNR}_{\text{LP}_t}$ in Fig.~\ref{fig:val} and choose those filter strengths which operate at this level. These values are marked with black circles in Fig.~\ref{fig:val} and are listed in Tab.~I. The differences against MCTF are also provided. 
	Accordingly, with $\text{WLDU}$ and a filter strength of $h{=}16{\cdot}\sigma_n^2$ bit rate savings of only $0.4758\%$ are achievable. For $\text{WLDUr}$ with a filter strength of $h{=}8{\cdot}\sigma_n^2$, coding gains of $1.9343\%$ compared to MCTF are possible. Considering $\text{WLDP}$ with a filter strength of $h{=}24{\cdot}\sigma_n^2$, bit rate savings of $3.0575\%$ compared to MCTF can be achieved. Finally, $\text{WLDPU}$ achieves coding gains of $4.3912\%$ compared to MCTF for a filter strength of $h{=}8{\cdot}\sigma_n^2$.
	To confirm that these averaged results are representative for all 327 sequences, we additionally determine the minimum and maximum achieved compression results for the specific noise parameters from Tab.~I. for all proposed setups. The results can be seen in Fig.~5, where the whole range of achieved bit rate savings per sequence is depicted.
	
\begin{figure*}
		\centering
		\begin{tikzpicture}  
		\node[align=center] (ref) at (0,0) {\includegraphics[width=0.25\textwidth]{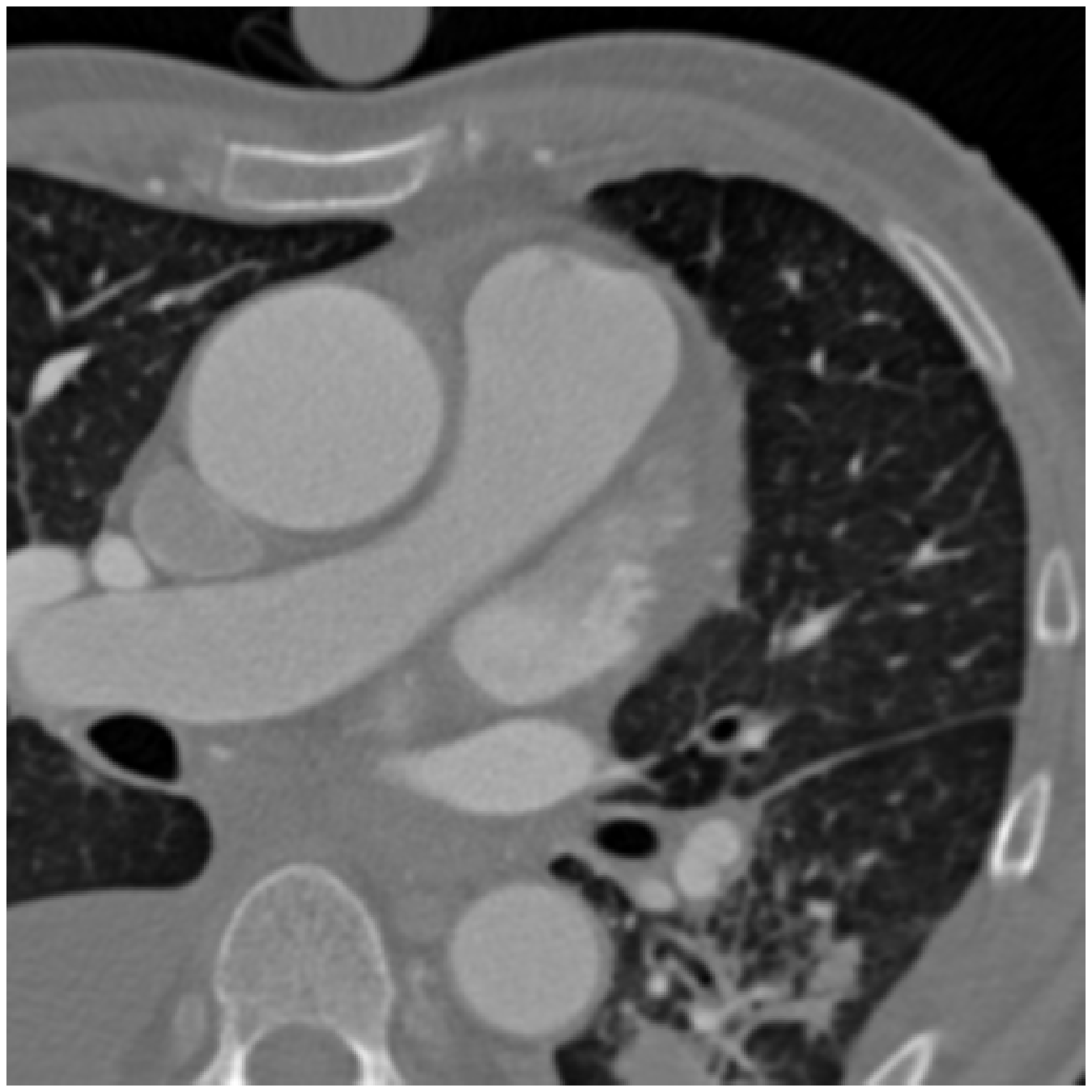}};
		\node[align=center](ref_label) at (0,-1.8) {Reference frame $f_1$};
		\draw[draw=red]  (0.1,0.75) rectangle (0.1+0.4,0.75+0.3);
		\draw[draw=red]  (-0.7,-0.55) rectangle (-0.7+0.4,-0.55+0.3);
		\draw[draw=red]  (-0.6,0.55) rectangle (-0.6+0.4,0.55+0.4);
		\node[align=center] (cur) at (4.5,0) {\includegraphics[width=0.25\textwidth]{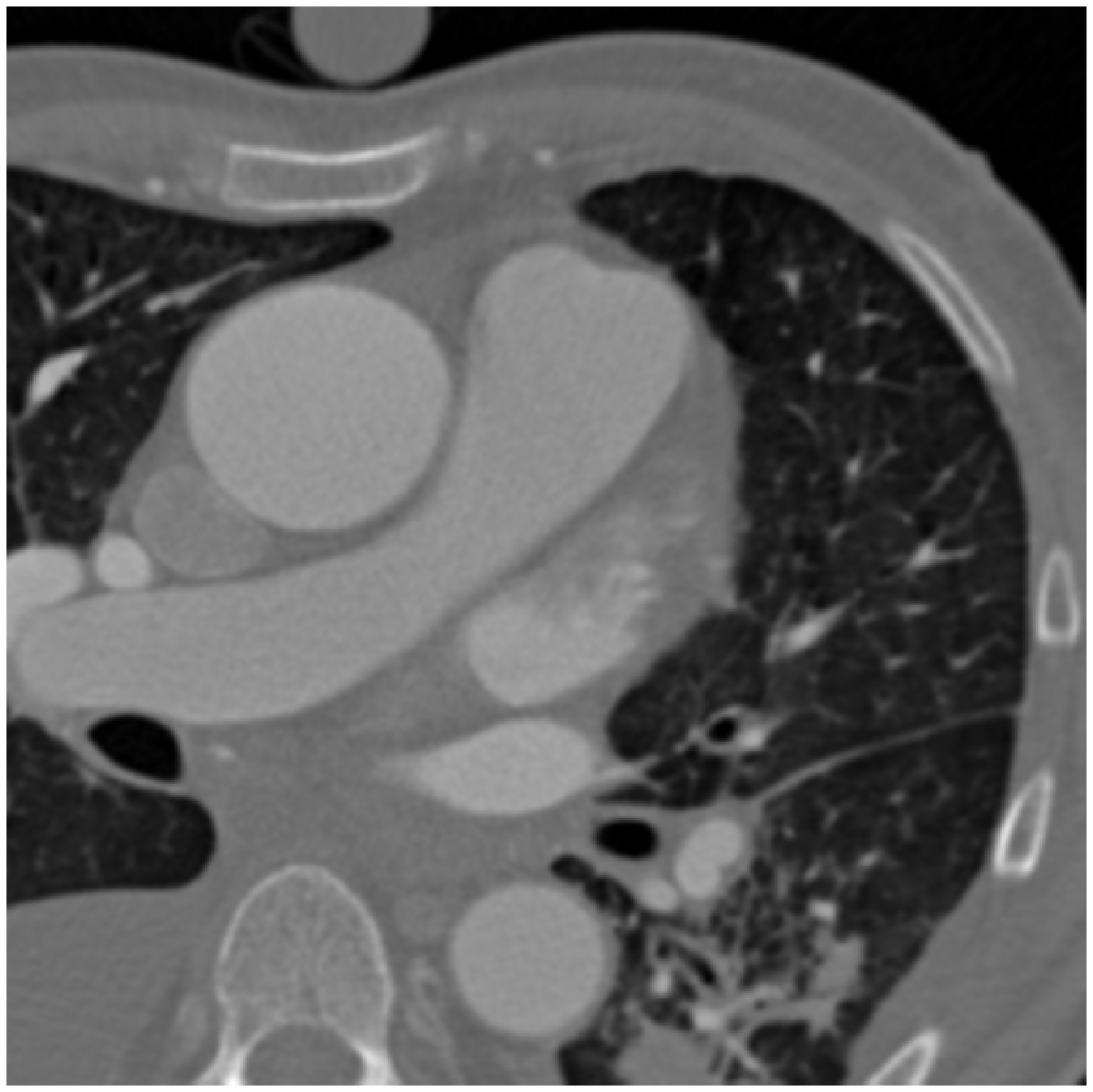}};
		\node[align=center](cur_label) at (4.5,-1.8) {Current frame $f_2$};
		\draw[draw=red]  (4.6,0.75) rectangle (4.6+0.4,0.75+0.3);
		\draw[draw=red]  (3.8,-0.55) rectangle (3.8+0.4,-0.55+0.3);
		\draw[draw=red]  (3.9,0.55) rectangle (3.9+0.4,0.55+0.4);
		\node[align=center] (mctf) at (9,0) {\includegraphics[width=0.25\textwidth]{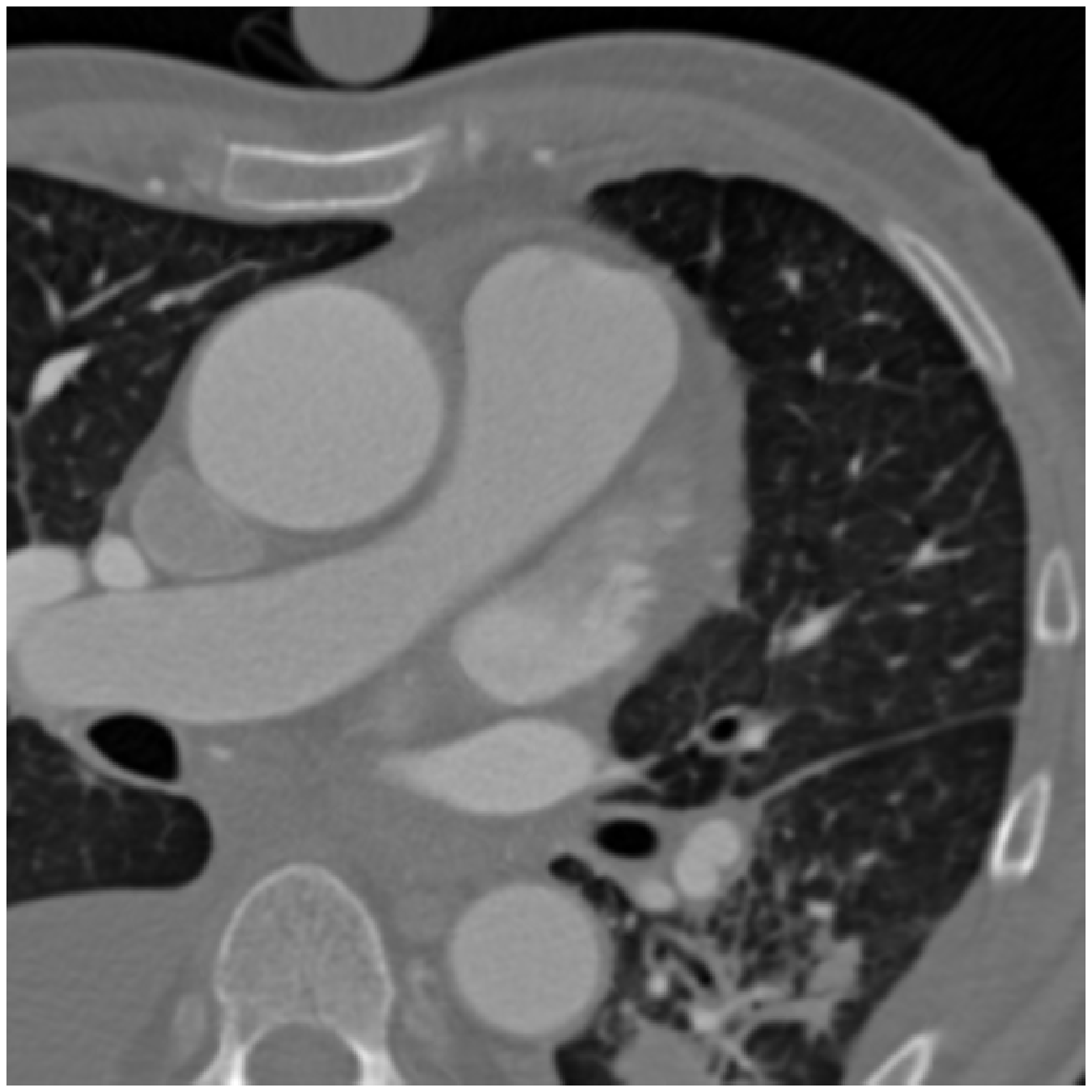}};
		\node[align=center](mctf_label) at (9,-1.8) {$\text{LP}_1$ from MCTF};
		\node[fill=white,position=-54:{-12mm} from mctf] (n1) {\small{$49.20$\,dB}};
		\node[fill=white,position=234:{-12mm} from mctf] (n2) {\small{$303.21$\,kB}};
		\node[align=center] (wldpu) at (13.5,0) {\includegraphics[width=0.25\textwidth]{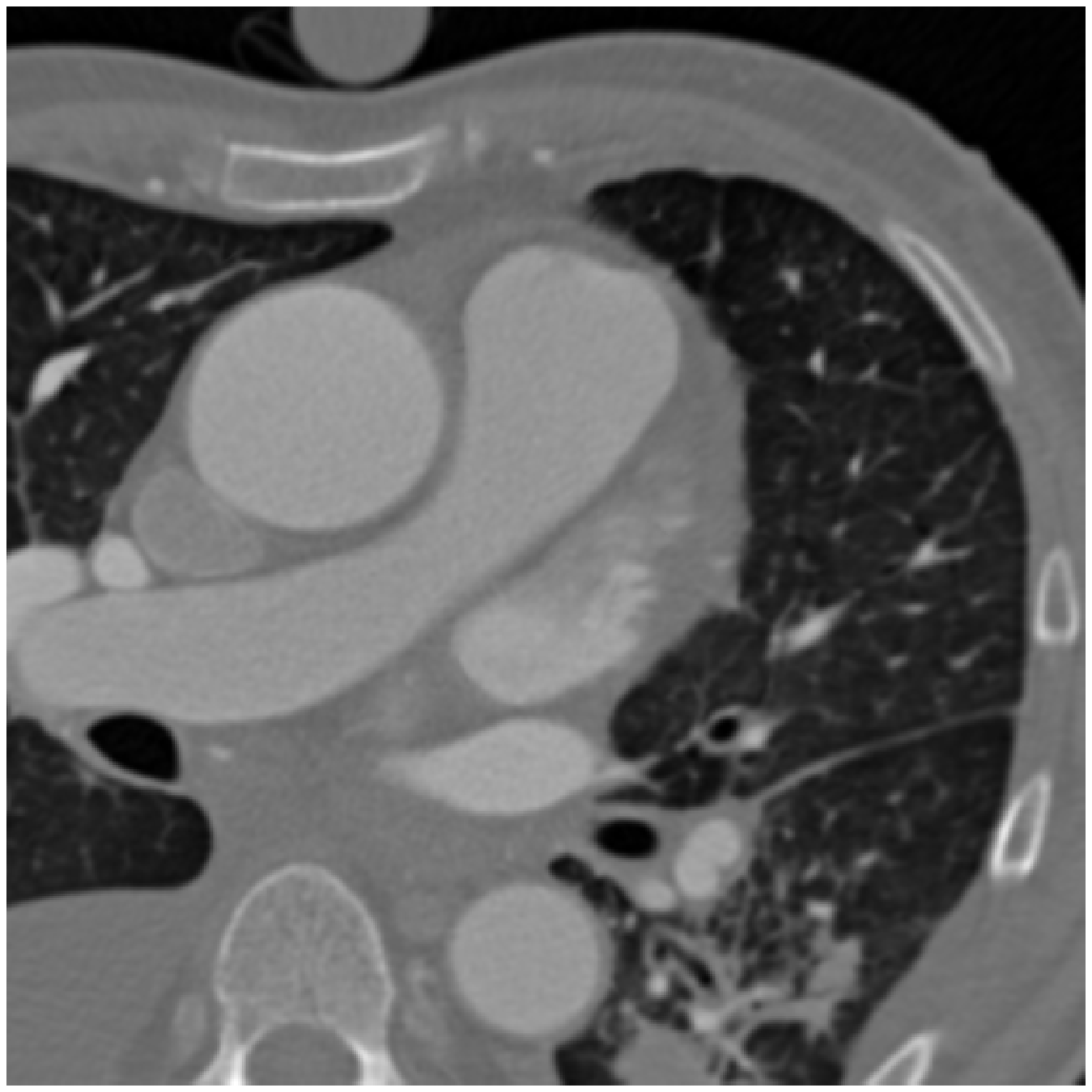}};
		\node[align=center](wldpu_label) at (13.5,-1.8) {$\text{LP}_1$ from WLDPU};
		\node[fill=white,position=-54:{-12mm} from wldpu] (n3) {\small{$49.10$\,dB}};
		\node[fill=white,position=234:{-12mm} from wldpu] (n4) {\small{$292.84$\,kB}};
		\end{tikzpicture}	
		\caption{Visual results comparing LP frames from WLDPU and MCTF against original frames. Dominant moving areas are highlighted. $\text{PSNR}_{\text{LP}_t}$ and file size to reconstruct $f_1$ and $f_2$, comprising $\text{LP}_1$, $\text{HP}_1$, and the corresponding MVs, are provided.}
		\label{fig:visual_results}
		\vspace{-3ex}
	\end{figure*}

	In Fig.~\ref{fig:visual_results}, some visual results are provided. From left to right, the reference frame $f_1$, the current frame $f_2$, $\text{LP}_1$ for MCTF, and $\text{LP}_1$ for WLDPU from data set $1$ for slice position $z{=}1$ can be seen.
	In the original frames, some dominant moving areas are marked by red boxes. As can be seen, both LP frames provide very similar visual results, which is also confirmed by the $\text{PSNR}_{\text{LP}_t}$ values. Additionally, the file size to reconstruct $f_1$ and $f_2$ at decoder side, including the file size of $\text{LP}_1$, $\text{HP}_1$, and the corresponding MVs, is given. This rate is reduced by $3.42\%$ from MCTF to WLDPU.   
	
	\subsection{Comparing different filters}
	
	\begin{figure}[t]
		\centering
		\resizebox{0.47\textwidth}{!}{%
			\input{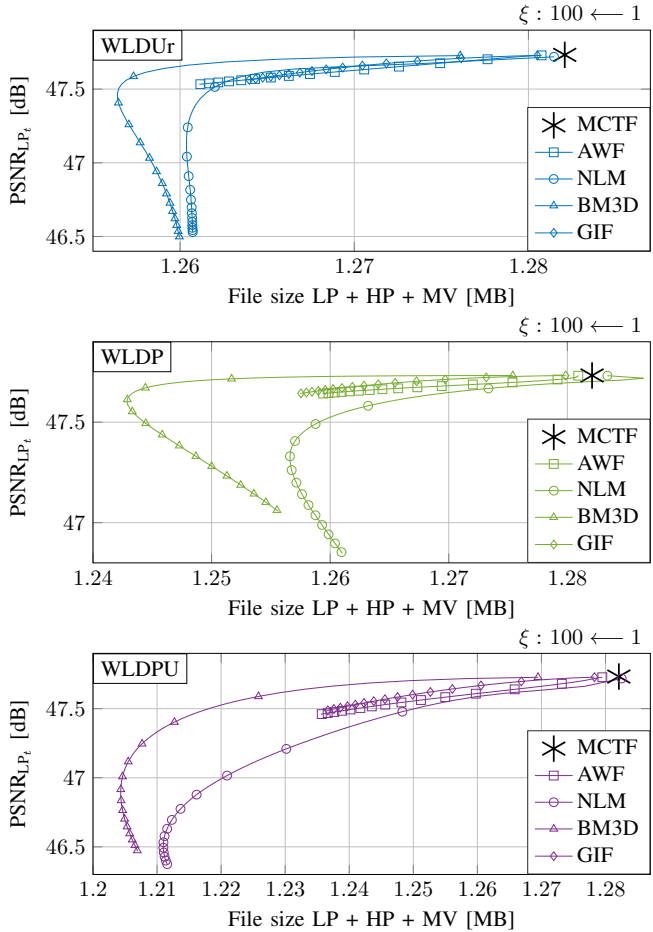}}
		\caption{Comparing different filters for WLDUr, WLDP, and WLDPU. Results are averaged over all 327 sequences.
		}
		\label{fig:WLDUr}
		\vspace{-3ex}
	\end{figure}	
		
	As a second experiment, different filters for each setup are evaluated. In addition to BM3D filtering, the Adaptive Wiener Filter\,(AWF)~\cite{1990ph...book.....L}, the Non-Local Means algorithm\,(NLM)~\cite{1467423}, and Guided Image Filtering\,(GIF)~\cite{He2010} is applied.
	These filters were already applied in~\cite{Lanz2018} to evaluate WLDU. 
	The upper diagram of Fig.~\ref{fig:WLDUr} shows the performance of all filters and filter strengths for WLDUr. For AWF and GIF, the quality of the LP subband is kept at a very high level for the whole range of tested filter strengths, while the overall file size is reduced significantly. In contrast, the NLM and BM3D filters show this behavior only until a filter strength of $h{=}8{\cdot}\sigma_n^2$, afterwards the quality drops very fast and the rate increases. However, the achieved compression ratio of BM3D for this filter strength is outstanding compared to all other filter setups.
	In the middle diagram of Fig.~\ref{fig:WLDUr}, the performance of all filters and filter strengths for WLDP is given. The same behavior for all filters like for WLDUr can be observed, but in a larger stretch. This means, further bit rate savings compared to WLDUr are achieved at approximately the same level of $\text{PSNR}_{\text{LP}_t}$. Again, the BM3D filter performs best.
	The performance of all filters and filter strengths for WLDPU can be observed in the lower diagram of Fig.~\ref{fig:WLDUr}. It shows that all filters profit by the combination of WLDP and WLDUr. Consistently, BM3D filtering achieves the best results.
			
	Obviously, the right choice of the filter strength $h$ is very important for achieving a high compression ratio and keeping the quality of the LP subband in terms of $\text{PSNR}_{\text{LP}_t}$ at a high level.
	Therefore, the next step of our work targets the generation of a suitable signal model, which allows for a theoretical analysis of the noise impact to the entire framework as well as an optimum choice of the filter strength by minimizing $\sigma_n^2$ in both subbands.
	Nevertheless, we showed that it is possible to outperform state-of-the-art methods like WLDU and the truncated WT with our proposed modifications. This is very advantageous for professional applications, where a scalable representation always requires a suitable trade-off regarding the quality of the LP subband and the corresponding compression ratio.
	
	\section{Conclusion}
	\label{sec:conclusion}
	A scalable representation of video data is often required in professional applications. This can efficiently be reached by wavelet-based approaches. By incorporating MC methods into the lifting structure, the quality of the LP subband, which serves as a temporally downscaled representative of the original sequence, can be increased significantly. However, the compression efficiency of dynamic CT data suffers enormously from MCTF. By integrating denoising filters into the lifting structure, this drawback can be lessened. In this work, a novel processing order for an already existing approach of denoising in the update step was introduced. Further, a second denoising filter in the prediction step was proposed. Both modifications offer a significant enhancement of the compression efficiency for dynamic CT data, while the quality of the LP subband is kept nearly constant for a certain range of filter strengths. Additionally, the property of lossless reconstruction is still given, which makes our novel approach very attractive for high-fidelity application scenarios, where the channel capacity is limited.

	\bibliographystyle{IEEEbib}
	\bibliography{Literatur.bib}

\end{document}